\documentstyle[11pt,newpasp,twoside,epsf]{article}
\begin{document}

\title{Velocities of RR Lyrae Stars in the Sagittarius Tidal Stream}

\author{A. Katherina Vivas}
\affil{Centro de Investigaciones de Astronom{\'\i}a (CIDA), Apartado
Postal 264, M\'erida 5101-A, Venezuela}

\author{Robert Zinn}
\affil{Department of Astronomy, Yale University,  PO Box 208101. New Haven, 
CT 06520-8101, USA}

\author{Carme Gallart}
\affil{Instituto de Astrof{\'\i}sica de Canarias (IAC), Calle V{\'\i}a 
L\'actea, E-38200 La Laguna, Tenerife, Canary Islands, Spain}

\begin{abstract}
We have measured radial velocities and metallicities of 16 RR Lyrae
stars, from the QUEST survey, in the Sagittarius tidal stream at 50
kpc from the galactic center.  The distribution of velocities is quite
narrow ($\sigma=25$ km/s) indicating that the structure is coherent
also in velocity space. The mean heliocentric velocity in this part of
the stream is 32 km/s.  The mean metallicity of the RR Lyrae stars is
[Fe/H]$=-1.7$. Both results are consistent with previous studies of
red giant stars in this part of the stream. The velocities also agree
with a theoretical model of the disruption of the Sagittarius galaxy.
\end{abstract}

\section{Introduction}
Numerous observations have shown that the Sagittarius dwarf spheroidal
galaxy (Sgr) is being disrupted by the tidal forces of the Milky
Way. A long stream of its tidal debris has been observed multiple
times in different parts of the sky. Many of these observations are
described elsewhere in these proceedings. They include an all-sky view
of M giant stars (Majewski et al. 2003), RR Lyrae stars (Vivas et
al. 2001, Ivezic et al. 2000), A stars (Yanny et al. 2000), halo
turnoff stars (Newberg et al. 2002) and main-sequence stars in 
color-magnitude diagrams (Mart{\'\i}nez-Delgado et al. 2001). Each of these
observations detects the stream as an over-density of the tracer above
the halo background. Simulations of the disruption of satellite
galaxies by the Milky Way show that tidal streams should be seen not
only as over-densities but also as coherent structures in velocity
space (e.g., Harding et al. 2001).

\begin{figure}
\plottwo{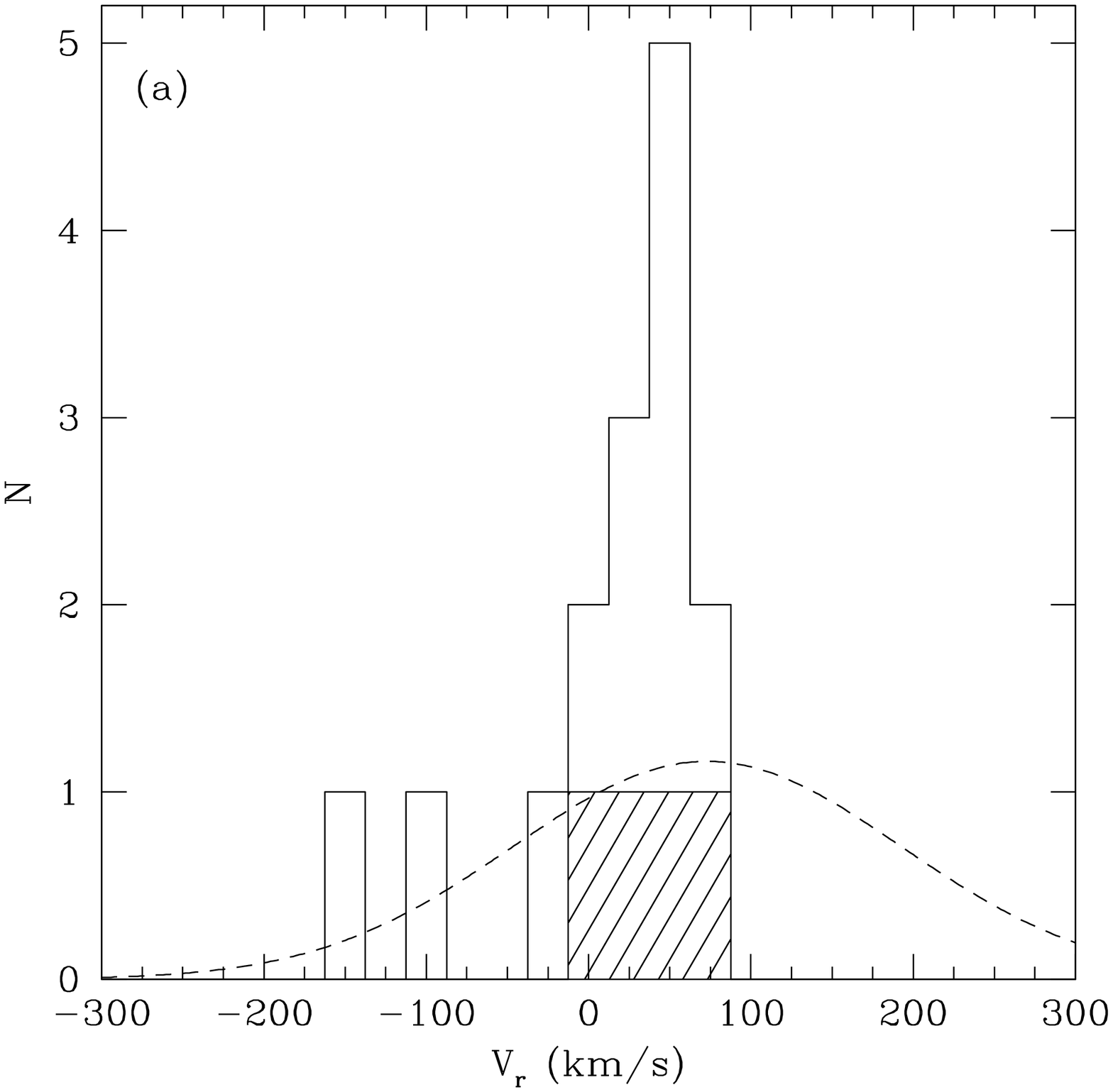}{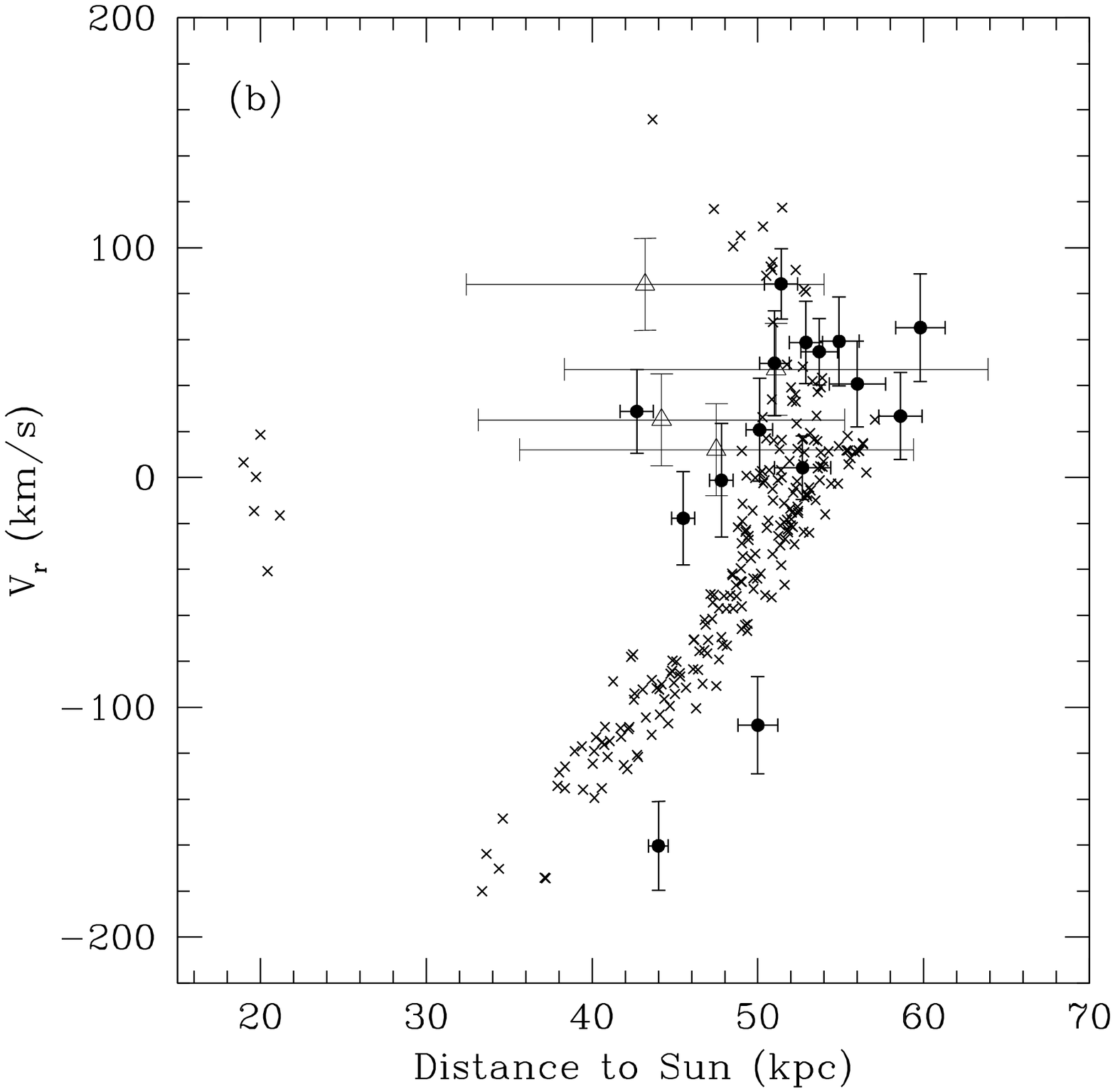}
\caption{(a) Histogram of the heliocentric radial velocities of 16
RR Lyrae stars observed with VLT. For comparison we show the distribution
of velocities of red giant stars from the Spaghetti survey (shaded histogram).
The dashed line shows the expected distribution of velocities of a sample
of random halo stars. (b) Velocities of the RR Lyrae stars (solid circles)
as a function of distance to the Sun. Our measurements agree with the
predictions of the theoretical models of Mart{\'\i}nez-Delgado et al. (2003),
shown as crosses.
Open triangles correspond to the red giants from the Spaghetti survey.}
\label{fig-vel}
\end{figure}

The QUEST survey for RR Lyrae stars (Vivas et al. 2001, 2003, see also Zinn 
et al. in this volume) has observed part of the
Sgr stream in a long, $2.3^\circ$-wide strip near the celestial 
equator. We present here a study of the radial velocities of a sub-sample 
of 16 RR Lyrae stars in the Sgr tidal stream. RR Lyrae stars stand out
as one of the best tracers of the old halo stellar population because they are
bright standard candles. Thus, they can provide excellent views of the
stream in both the three-dimensional spatial distribution and the radial 
velocity distribution.

\section{The Data}

The 16 RR Lyrae stars belong to the clump located at $\sim50$ kpc from
the galactic center which has been related to the leading arm of the
Sgr tidal stream.  The QUEST survey found 84 stars in the Sgr stream,
a factor of 10 higher than the background of halo stars.  The spatial
distribution of the clump indicates that is quite wide in right
ascension, about $36^\circ$, from $13\fh 0$ to $15\fh 4$. We included
in this study stars along all the stream in order to confirm its true
size.  All stars have mean magnitudes of $V \sim 19.2$.

\begin{figure}
\plotfiddle{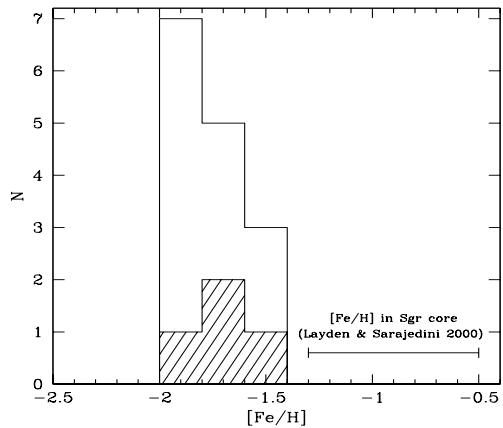}{5.5cm}{0}{35}{30}{-110}{-50}
\caption{Histogram of the distribution of [Fe/H] of 16 RR Lyrae stars in
the Sgr stream. For comparison we show the abundances of the red giants
from the Spaghetti survey (shaded region). We also indicate the range
of metallicities observed in the core of the Sgr galaxy.}
\end{figure}

Because RR Lyrae stars are pulsating stars with periods of $\sim0.5$
days, exposure times of spectra should be kept short ($\la 30$ min) in
order to avoid excessive broadening of the spectral lines by the
changing pulsational velocity.  Given the faintness of the stars in
the clump, a large telescope was needed.  Spectra of the 16 stars were
taken with FORS2 at the VLT-Yepun in Paranal, Chile, during June-Aug
2002. We used grating 600B which gives a resolution of $\sim 6$\AA,
and covers a spectral range from 3400-6300\AA. 
Exposure times varied between 20 and 30 minutes. 
For each star we obtained two spectra taken
at random times on different nights. This allowed us to make
measurements at two different phases during the pulsation cycle. A few
radial velocity standards were also observed with the same
instrumental setup.

\section{Radial Velocities}
Radial velocities of RR Lyrae stars change during the pulsation cycle
by up to $\sim 100$ km/s. Thus, it is important to know the exact
phase at which the spectrum was taken in order to separate the
systemic velocity of the star from the velocity due to the
pulsations. We obtained phase information from the QUEST RR Lyrae
catalog (Vivas et al 2003) which provides accurate ephemerides for all
the stars in our sample.  
We determined the radial velocities by cross-correlation with each of
the observed radial velocity standards. The error in a single
measurement is estimated to be $\sim 20$ km/s.
For each star we fitted a radial velocity
curve template following the procedure described in Layden (1994).  In
a few cases, the spectra was taken near the phase of maximum
brightness of the light curve. We did not use these observations since
there is a strong discontinuity in the radial velocity curve at this
point.

The results are shown in Figure 1a. The histogram shows the
distribution of the heliocentric radial velocities of the 16 stars.
Taking out the two obvious outliers, the distribution is quite narrow,
with a mean of 32 km/s and a standard deviation of only 25 km/s.  The
distribution does not resemble the one expected for a random sample of
halo stars, which is the dashed, Gaussian curve in Figure
1a.  For comparison we also show the distribution of velocities of
four red giant stars from the Spaghetti survey (Dohm-Palmer, et
al. 2001) which seem to be also associated with the Sgr stream in a region
of the sky very close to ours. The presence of two outliers is not
surprising.  If the Sgr stream lies within a smooth distribution of halo
stars following a $r^{-3}$ power-law, we expect 1-2 halo RR
Lyrae stars in this volume of the sky.

There is also good agreement between our observations and the
predictions of the models of the disruption of Sgr. We compare with
one of these models (Mart{\'\i}nez-Delgado et al 2003) in Figure 1b. The
red giants from the Spaghetti survey are also included.

\section{Metal Abundances}
The metal abundances of the stars were measured using the modified
$\Delta S$ technique described by Layden (1994), which is based on the
equivalent widths of the Ca II K line and the Balmer lines.  The error
in a single measurement of [Fe/H] is 0.2 dex.  The distribution of
metallicities of the 16 stars of our sample is shown in Fig 2. Our
results are in very good agreement with the 4 red giants from the
Spaghetti survey.  The mean metallicity of the RR Lyrae stars is
[Fe/H]$= -1.7$.  Notice however that the mean metallicity of the
stream is significantly lower than in the core of the Sgr galaxy. This
could be explained if Sgr once had a radial gradient in metallicity,
an age-metallicity relation (since the RR Lyrae variables are
exclusively very old stars) or a combination of both.

\section{Conclusions}

We have measured VLT spectra for 16 RR Lyrae stars belonging to a part
of the Sgr tidal stream, located at 50 kpc from the galactic center.
The distribution of radial velocities is quite narrow, indicating that
the Sgr clump is a coherent structure in velocity space.  We do not
find significant gradients of radial velocities or metallicities along
the stream.

\smallskip

\acknowledgments
This work is based on observations collected at the European Southern 
Observatory, Chile. The data were obtained as part of an ESO Service Mode
run. This research project was partially supported by the National Science 
Foundation under grant AST-0098428.

\end{document}